\documentclass[aps,prl,twocolumn,superscriptaddress,groupedaddress]{revtex4}  
\usepackage{graphicx}  
\usepackage{dcolumn}   
\usepackage{bm}        
\usepackage[T1]{fontenc}
\usepackage[utf8]{inputenc}
\usepackage{amssymb}   
\usepackage{amsmath}

\usepackage{float}

\begin{document}

\title{Crowding competes with trapping to enhance interfacial diffusion}
\author{Mislav Cvitkovi{\'c}} 
\thanks{These authors contributed equally}
\affiliation{Institute of Ruđer Bo{\v s}kovi{\'c}, Division of Physical Chemistry, Group for Computational Life Sciecens, Bijeni{\v c}ka c. 54, Zagreb 10000, Croatia}
\affiliation{Friedrich-Alexander University Erlangen-N\"urnberg, Institute for Theoretical Physics, PULS Group, IZNF, Cauerstrasse 3, Erlangen 91058, Germany} 
\author{Dipanwita Ghanti}
\thanks{These authors contributed equally}
\affiliation{Institute of Ruđer Bo{\v s}kovi{\'c}, Division of Physical Chemistry, Group for Computational Life Sciecens, Bijeni{\v c}ka c. 54, Zagreb 10000, Croatia}
\author{Niklas Raake}
\affiliation{Friedrich-Alexander University Erlangen-N\"urnberg, Institute for Theoretical Physics, PULS Group, IZNF, Cauerstrasse 3, Erlangen 91058, Germany} 
\author{Ana-Sun{\v c}ana Smith{\footnote{Corresponding author; e-mail: smith@physik.fau.de, asmith@irb.hr}}}
\affiliation{Institute of Ruđer Bo{\v s}kovi{\'c}, Division of Physical Chemistry, Group for Computational Life Sciecens, Bijeni{\v c}ka c. 54, Zagreb 10000, Croatia} 
\affiliation{Friedrich-Alexander University Erlangen-N\"urnberg, Institute for Theoretical Physics, PULS Group, IZNF, Cauerstrasse 3, Erlangen 91058, Germany} 

\date{\today}

\begin{abstract}
Diffusion in the crowded environments of the biological membranes or materials interfaces often involves intermittent binding to surface proteins or defects. To account for this situation we study a 2-dimensional lattice gas in a field of immobilized traps. Using kinetic Monte Carlo simulations, we calculate the effective diffusion coefficient in the long-time limit as a function of the traps and particle densities. We find a remarkable  result - an increase of the diffusion coefficient with particle density, an effect that we coin as \textit{crowding-enhanced diffusion}. We rationalize this result using scaling arguments and the master equation approach.

\end{abstract}

\maketitle


Diffusive transport at interfaces is an ubiquitous process in nature. Prototypical examples involve proteins on a phospholipid membrane of a living cell \cite{franosh13}, nanoparticles in liquid-filled solid pores \cite{alvaro16}, or molecules interacting with growing or functional materials interfaces \cite{kumar18}. The characteristic of this transport is that it takes place on a two-dimensional surface,  which by its structure and molecular composition may be very complex. Often transport takes place at an appreciable concentration of diffusing species which furthermore interact with defects on the surface or specifically incorporated functional moieties such as proteins on the membranes \cite{smith18}. These local interactions trap the diffusing species, intermittently arresting the particle, which in turn reflects on the effective diffusion coefficient.  

Diffusion on interfaces has been theoretically studied as a simplified problem of a stochastic motion on a lattice \cite{saxton87,saxton94}. In these models, steric hindrance between diffusing particles is typically introduced, which prevents double occupancy of the same lattice site. For a gas of diffusing particles in these conditions, long-time diffusion coefficient has been determined by calculating many-particle correlation functions \cite{nakazato80,tahir83,beijeren85}, as well as by considering memory effects of diffusing particles \cite{halpern96}. These different approximations were recently complemented with the exact calculation of the tracer particles' probability distribution function on a crowded lattice \cite{pigeon17}, yielding the diffusion coefficient as a function of gas density or surface coverage. 
\begin{figure}
\includegraphics[scale=0.4]{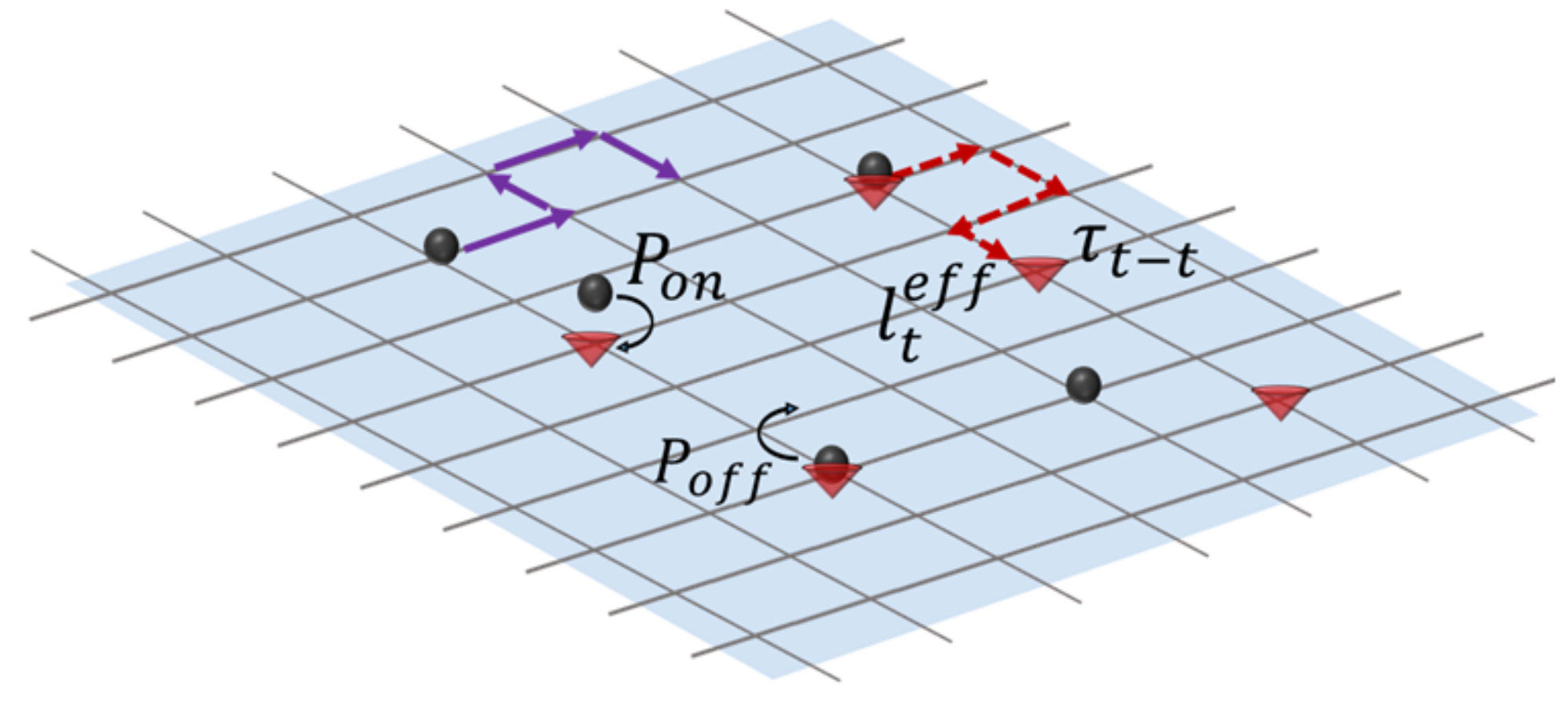}
\caption{ Depiction of a 2-D lattice gas (solid spheres) in the field of traps (red cones).  If the particle and trap position coincide, then the particle can bind to the trap with a probability $ P_{on} $ and unbind with a probability $ P_{off}$. Length and time traversed by the particle during diffusion from one trap to another is referred as $ l_{t}^{eff} $ and $ \tau_{t-t} $, respectively, while the time in the trap is denoted as $\tau_{off}$}. 
\label{fig_model}
\end{figure}

While the increased density of diffusing species rescales the diffusion constant at long times \cite{dix08}, interactions with defects, however, yield a rich dynamic behavior\cite{jeon11,jeon16}, which was extensively studied using theoretical \cite{sokolov12,yann18,metzler19} and computational means \cite{evers13,samanta16,schnyder18}. Particular efforts were focused on understanding the effect of interaction types on the diffusion of a tracer particle, which lead to the random trap model \cite{bouchaud90}, Havlin–Weiss comb model \cite{weiss86}, or the quenched trap model \cite{burov07}. Computational approaches provided insight on the effects of surface heterogeneity using the kinetic lattice gas model \cite{mak88}, or the bivariate trap model \cite{viljoen96}. Furthermore, kinetic Monte-Carlo simulations were used to study the diffusion of a tracer on a surface with binding sites characterized by multiple energy levels \cite{saxton96}. It has been established that the tracer diffusion in the field of traps exhibits a sub-diffusive motion on short time scales and normal diffusion on long time scales \cite{franosh13,saxton96,saxton07,kusumi05}. These works provided significant impact on the role of the trap energetics on the diffusivity of a tracer \cite{bouchaud90}, however, the combined effects of crowding and trapping were not addressed so far in full depth to our knowledge. 

In this letter, we aim to rectify this situation by providing an approximate theory for surface diffusion of crowded particles in a field of simple traps.  We propose an expression for the long-time effective diffusion coefficient of the crowding particles using the scaling argument as well as the master equation approach. The model is validated by a favorable comparison with Monte Carlo (MC) simulations \cite{bihr15} (methodological details in Supplementary Information). In agreement with previous works \cite{franosh13,saxton96,saxton07,kusumi05}, we find that an intermittent sub-diffusive regime is inherited to the system as soon as the traps are introduced. More surprisingly, however, we find that in the long time limit the diffusion coefficient may be enhanced by crowding effects for any concentration of traps, at significant range of densities of lattice gas.


We start with placing $ N_{w} $  particles, mutually interacting with a hard wall repulsive potential, on a 2D  square lattice with $ N_{g} $  lattice sites of a length $ a_{0} $. The concentration of gas particles on the lattice is $ c_{w}=N_{w}/N_{g} $.  A random walking particle will traverse a distance $ a_{0} $ during time $ \tau=a_{0}^{2}/4D_{0} $, where $ D_{0} $ is the diffusion coefficient of a particle at infinite dilution. The concentration-dependent diffusion coefficient $ D_{cr}$ is determined as $ D_{cr}(c_{w},D_{0})=D_{0}p(c_{w}) $ where $ p(c_{w}) $ is the concentration-dependent  probability of a jump, whereby the larger the concentration, the smaller the effective diffusion coefficient.
 $p(c_{w})$ can also be seen as the factor normalizing the characteristic time to make a step in the crowded environment $\tau_{cr}=\tau/p$. It has been calculated using different approaches,  \cite{nakazato80,tahir83,beijeren85}, while here we estimate it from backward correlations as in the anti-persistent random walk (APRW) model \cite{halpern96}. The appropriateness of this choice, which is a compromise between accuracy over the entire density range and simplicity, as demonstrated by comparison with MC simulations and with other models (see Supplementary Information, section I).

The lattice is furthermore decorated by $ N_{t} $ randomly placed traps, the concentration of which is denoted as $c_{t}= N_{t}/N_{g} $. Upon hopping onto a site with a trap, a diffusing particle binds with the probability $ P_{on} $ ( Fig.\ref{fig_model}), and unbinds with the probability $P_{off}$.  Following the detailed balance condition, the binding energy is $ \Delta E_{b}=-k_{B}T ln(P_{on}/P_{off})$. The latter sets the concentration of bound particles $\langle c_{b}\rangle $, which can be calculated analytically from the partition function of the system, and its exact form is given in the SI \cite{mislav17}.

To estimate the diffusion constant in the long time limit  for the lattice gas in the field of traps $D_{eff}$, we aim at coarsening the diffusion process and finding the scaling function $f$ 
\begin{equation}
D_{eff}(c_{w},c_{t},P_{on},P_{off})=f(c_{w},c_{t},P_{on},P_{off})D_{cr}(c_{w}).
\label{eq-D_eff}
\end{equation}
Specifically, we presume that the effective diffusion coefficient can be related to the square of the average distance between two traps at which binding actually occurs $ \langle  l_{t}^{2}\rangle ^{eff}$ , and  the time it takes to make this coarsened diffusion step $\tau^{eff}$, i.e.  $4D_{eff} = \langle   l_{t}^{2}\rangle ^{eff}/\tau^{eff}$.  
Here, the average square distance between two efficient traps can be related to the effective concentration of traps  $\rho^{eff}$ as $ \langle l_{t}^{2} \rangle =a_{0}^{2}/\rho^{eff}$. The latter can be estimated from the density of free traps $c_{t}-\langle c_{b}\rangle $, to which a diffusing particle can bind with the probability $P_{on}$ such that  $ \rho^{eff}=(c_{t}-\langle c_{b} \rangle )P_{on}$. 

\begin{figure}. 
\includegraphics[scale=0.43]{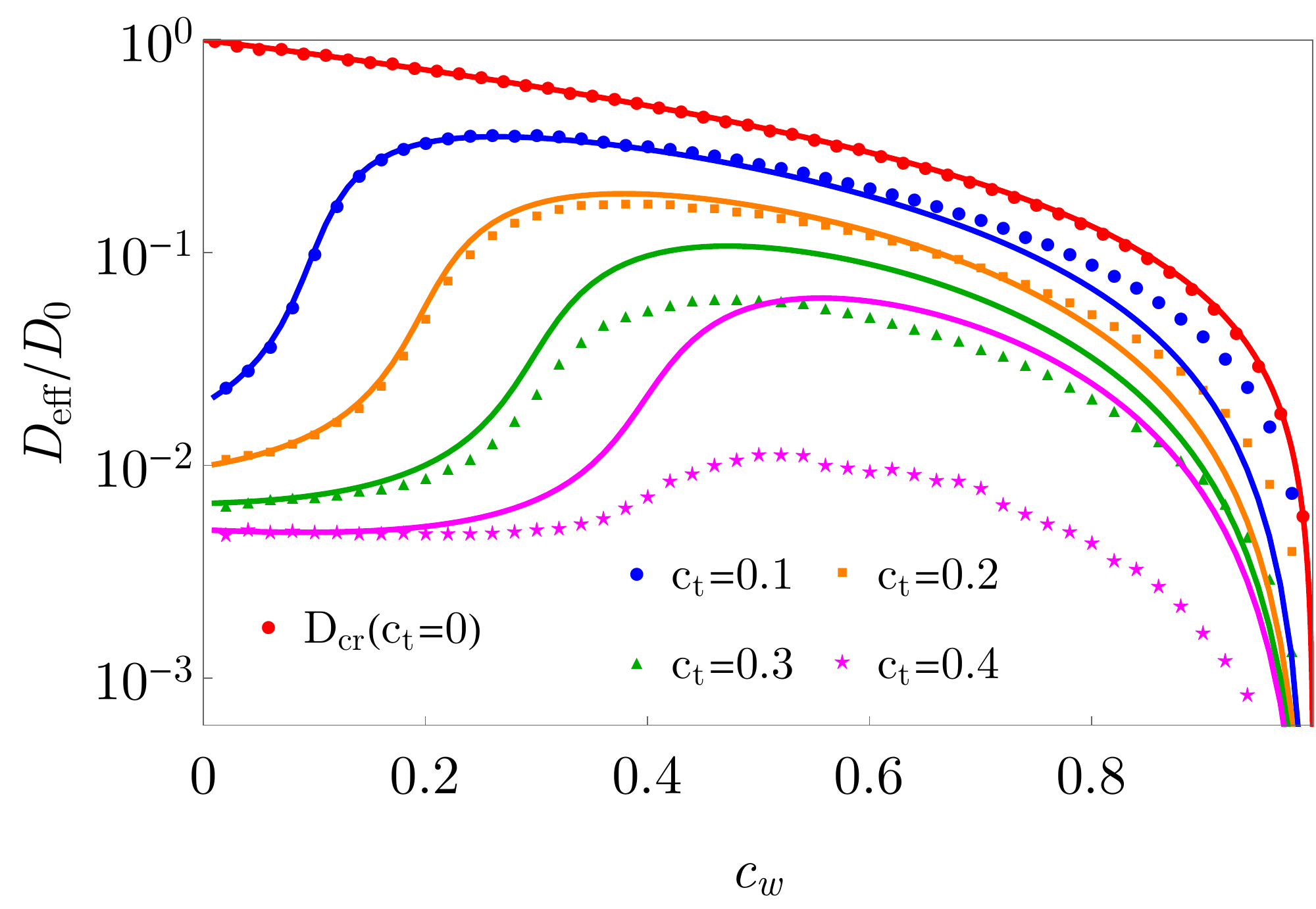}
\caption{ Effective diffusion coefficient, normalized by the diffusion coefficient at infinite dilution,  as a function of the lattice gas concentration $ c_{w} $ for various trap densities $c_t$ ( $ c_{t}=0, 0.1,0.2,0.3 $ and $ 0.4 $) with $ P_{on}=0.5 $ and $ P_{off}=0.001 $. Results of MC simulations are shown with symbols, and theory (eq.(\ref{eq-D_eff2})) with lines.}
\label{fig_Deff}
\end{figure} 

The characteristic time $\tau^{eff}$ comprises of the time that it takes to leave a bounded trap $ \tau_{off} $, and the time to diffuse to the next trap where it binds again $ \tau_{t-t} $, which is simply $\tau_{t-t}=\langle l_{t}^{2}\rangle ^{eff}/4D_{cr}=\tau_{cr}/\rho^{eff} $. To leave the trap, a particle must unbind  with the probability $ P_{off} $ and make a step with the probability $ p(c_{w}) $, and hence, $ \tau_{off}=\tau_{cr} /[P_{off}p(c_{w})]$. With $\tau^{eff}= \tau_{off}+\tau_{t-t}$, it is straightforward to estimate the effective diffusion coefficient as
\begin{equation}
D_{eff}=D_{cr}\bigg(1+\frac{P_{on}}{P_{off}}\frac{c_{t}-\langle c_{b}\rangle }{p(c_{w})}\bigg)^{-1}.
\label{eq-D_eff2}
\end{equation}

The same result can be derived by considering the master equation approach by calculating the probability of finding a particle $ P(r,t)$, at position $ r $ at time $ t $. Since the later emerges as a sum of the probability of finding a particle unbound on that site $ P_{cr}(r,t) $ and the probability of finding the particle trapped on the same site $ P_{tr}(r,t) $, we can write $ P(r,t)=P_{cr}(r,t)+P_{tr}(r,t) $. The time evolution of both probabilities is given by:
\begin{eqnarray}
\frac{\partial P_{cr}(r,t)}{\partial t}=\frac{\gamma}{2d}\sum_{i=1}^{d}(P_{cr}(r-e_{i},t)+P_{cr}(r+e_{i},t)\nonumber\\
-2P_{cr}(r,t))-k_{on}^{e}P_{cr}(r,t)+k_{off}^{e}P_{tr}(r,t) ,
\label{eq-master1}
\end{eqnarray}
and
\begin{equation}
\frac{\partial P_{tr}(r,t)}{\partial t}=k_{on}^{eff}P_{cr}(r,t)-k_{off}^{eff}P_{tr}(r,t).
\label{eq-master2}
\end{equation}
Here, $\gamma=p(c_{w})/\tau $ is the hopping rate of the particle and $ k_{on}^{eff} $, $ k_{off}^{eff} $ are the effective binding and unbinding rates of any particle.
By introducing Fourier transform $ S(\textbf{k},w)=1/\pi \int_{\Omega}dr\int_{0}^{\infty}dt P(r,t)e^{i(wt-\textbf{k}.r)} $ of the eq.(\ref{eq-master1}) and eq.(\ref{eq-master2}) we can calculate the relation between dynamic structure factors $ S_{cr}(\textbf{k},w) $ and $ S_{tr}(\textbf{k},w) $. By calculating real part of total Fourier transform $ Re(S(\textbf{k},w)) $ and using the relation
 $ D_{eff} = \lim_{w \to 0}(\lim_{\textbf{k} \to 0}\pi w^{2}/\textbf{k}^{2}Re(S(\textbf{k},w)))$ we obtain
\begin{equation}
D_{eff}=D_{cr}\frac{k_{off}^{eff}}{k_{on}^{eff}+k_{off}^{eff}}
\label{eq-D_effM}
\end{equation}
We note that for $ k_{on}^{eff}=0 $ the effective diffusion constant $ D_{eff} $ reduces to $ D_{cr} $.
 
A particle can only bind to a trap if the trap is empty. Hence, the effective binding rate is $ k_{on}^{eff}=P_{on}(c_{t}-\langle c_{b} \rangle )/\tau $. Similarly, effective unbinding rate of a bounded particle depends on the availability of a free site to jump, so $ k_{off}^{eff}=P_{off}p(c_{w})/\tau $. Substituting these expression for effective rates in eq.(\ref{eq-D_eff}) yields $ D_{eff} $ equal to that in eq.(\ref{eq-D_eff2}) (see SI for details of the calculation).

\begin{figure}
\includegraphics[scale=0.43]{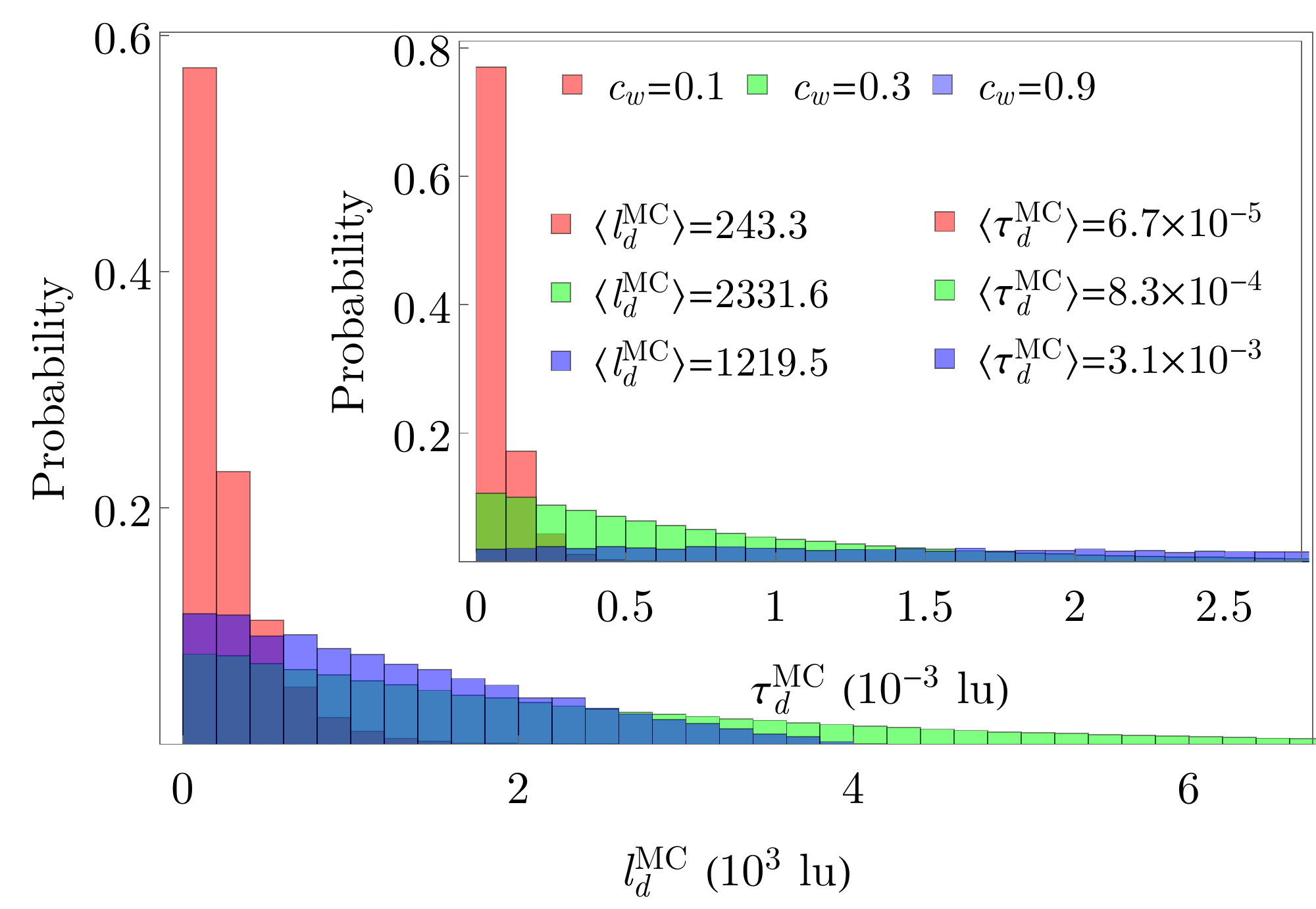}
\caption{
The probability distribution of the path length of a particle between two successive trapping events ($ l_{d}^{MC} $) is shown for $ c_{w}=0.1, 0.3 $ and $ 0.9 $ for fixed $ c_{t} = 0.1 $, $ P_{on}=0.5 $ and $ P_{off}=0.001 $ . In the inset, corresponding distribution of time between two successive trapping events of a particle ($ \tau_{d}^{MC} $) is shown.
Corresponding $\langle l_{d}^{MC}\rangle $ and $\langle \tau_{d}^{MC} \rangle $ is presented in the inset. The data is obtained from kinetic Monte-Carlo simulations as presented in lattice units.  
}
\label{fig_histogram}
\end{figure}

\begin{figure}
\includegraphics[scale=0.4]{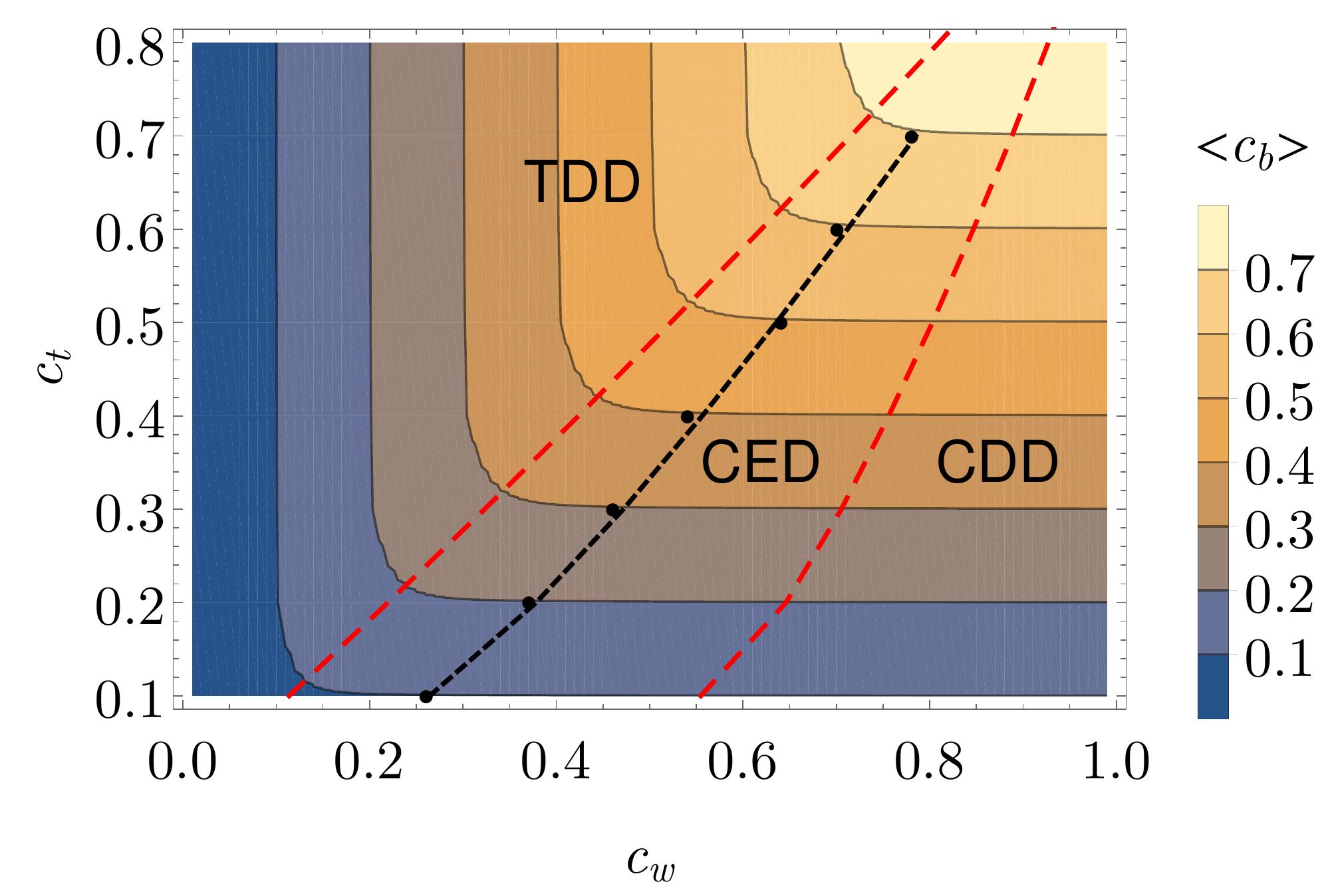}
\caption{Contour plot of the mean concentration of bound particles $ \langle c_{b}\rangle $  in the $ c_{w} $-$ c_{t} $ plane. The intersection points between solid black lines and red dashed lines represent the inflection points, calculated from our theory. Two red dashed inflection lines divide the plot into three regions, namely 
\textit{trapping-dominated diffusion} region (TDD), \textit{crowding enhanced diffusion} region (CED) and \textit{crowding-dominated diffusion} region (CDD). The black dashed line and points represent the critical concentration of crowding, at which the effective diffusion coefficient is maximized. The lines have been calculated from our theory, whereas the discrete data points have been obtained from MC simulation using $ P_{on}=0.5 $ and $ P_{off}=0.001 $.
}
\label{fig_cb}
\end{figure}
Intuitively, one expects a monotonic decay of the effective diffusion coefficient with increasing the concentration of particles $ c_{w}$, which indeed is the result observed in the absence of traps ($c_t=0$ (brown line and points in Fig.\ref{fig_Deff})). However, as soon as traps are introduced ($c_t>0$), the effective diffusion coefficient starts to non-monotonically vary with the gas density and a maximum in $D_{eff}(c_w)$ is predicted by eq.(\ref{eq-D_eff2}). This surprising property of lattice gas diffusion in the field of traps is confirmed by MC simulations (symbols in Fig. \ref{fig_Deff}). Particularly good agreement between theory and simulations is obtained at low $ c_{w} $ and $ c_{t} $. The strongest discrepancies occur for intermediate to high  $ c_{w} $ and $ c_{t} $, which suggest that higher-order correlations play an important role in this range of parameters.

In order to understand the underlying mechanism of for this behaviour, we extract the probability distribution of the actual path length traversed by the particle between two successive trapping events  $l_{d}^{MC}$ (Fig.\ref{fig_histogram}), and the distribution of time between two trapping events $\tau_{d}^{MC}$ (inset of Fig.\ref{fig_histogram}). We clearly see that at low concentrations of $c_{w}$, both typical lengths of the path, and the actual time between two trapping events is in average short, which means that the particles spend most time in the traps, this being detrimental to the effective diffusion coefficient. We denote this regime as "trapping-dominated diffusion".  For moderate $ c_{w}$, the characteristic $\tau_{d}^{MC}$ becomes significant while the tails of the distribution of $l_{d}^{MC}$ are the thickest. This means that at these concentrations the particles spend extended time meandering through the system. In this regime, a high level of trap occupancy is achieved by the significant concentration of walkers, but the crowding effects are not sufficient to prevent the diffusion - i.e. the walkers move away from occupied traps before they interact with a free trap. This yields "crowding enhanced diffusion". At high concentrations, nearly all traps are occupied, hence the walkers survive the longest between two trapping events, but they make significantly shorter paths than in the intermediate regime. This is because the likelihood for making a step onto a next site decreases significantly due to the high concentration of the particles, and therefore, there is enough time to actually interact of the trap. We denote this regime as "the crowding dominated diffusion".  

The three regimes are clearly denoted in the diffusivity phase diagram (Fig.\ref{fig_cb}) which highlights the importance of the mean number of occupied traps $ \langle c_{b} \rangle $. For a given $ c_{t} $, the latter increases with increasing $ c_{w} $  until saturation, which in the case of reasonable large binding affinities presumes that either nearly all walkers are bound, or that nearly all traps are occupied, depending on their relative total number.  In both cases, the particles are still diffusing and binding-unbinding kinetics are still ongoing. The boundaries between the three regimes are determined from the inflection points in $ D_{eff}/D_{0} $ vs $ c_{w} $ for a given $ c_{t} $ (red dashed lines in Fig.\ref{fig_cb}), with $ D_{eff}$ maximized for the particular concentration of particles $ c_{w} $ as determined analytically (black dotted line) and fully supported by simulations (black symbols).

In summary, we discussed the simultaneous effect of crowding and trapping on surface diffusion. Using scaling arguments and from the master equation approach, we show that the diffusion is directly related to the density of free traps and not to the absolute density of traps, the occupancy of which is defined by the density of the gas and affinity of the gas for the traps.  We find the so-called \textit{trapping-dominated diffusion} as long as there are more traps than walkers in the system allowing significant interactions of the two, and impeding the capacity of walkers to explore the system. When the number of traps and walkers are comparable \textit{crowding-enhanced diffusion} takes place. In this regime, a large number of traps are occupied but there is  a significant fraction of particles still able to diffuse in an environment that is not overly crowded, which optimizes the diffusion constant. Finally, if the number of walkers dominates, than crowding becomes significant even at relatively low $ c_{w} $. The traps are, by and large occupied, but the particles have a smaller likelihood to move to the next site as it is likely already occupied. Consequently the system displays \textit{crowding-dominated diffusion} where the effective diffusion constant continuously decays with the density of walkers. These results are confirmed by scaling arguments, analytical modeling and kMC simulations.

Recent theoretical studies on diffusion of rod-shaped active particles \cite{mandal20}, tracer diffusion inside active particles bath \cite{abbaspour21} and external force driven tracer diffusion \cite{illien18} diffusion is enhanced either by the energetics of active particles or by external force driven dynamics. However, in our minimalistic model of lattice gas in the presence of traps \textit{crowding-enhanced diffusion} arises solely because of the interplay between crowding and trapping as part of equilibrium thermodynamics. 

Our findings naturally still require a direct experimental confirmation. However, few recent experimental studies suggested that crowding can favorably affect diffusion. One example is the facilitated diffusion of DNA-recognizing protein during specific target search over a long DNA strand\cite{krepel16,brackley13,bauer13}, which is a system that shows similar features as our model. Non-monotonous behavior was also observed in simulations of a tracer diffusing in a field of particles crowding the environment, as a function of the density of crowder and the depth of the minimum of the tracer-particle interaction potential, which is a result consistent with our findings albeit in three dimensions \cite{putzel14}. In three dimension both experimental and theoretical studies on polymer transport in a crowded medium suggested that crowding enhances long-term diffusivity \cite{chien16,chien17}. It would be therefore interesting to extend our work from surface to volume diffusion and account for more complex behavior and properties of traps, a task that we plan to address in future.

We acknowledge the funding by ERC StG 2013–337283 of the European Research Council in the early stages of the project, which was later supported by the German Science Foundation program SFB 1411 Design of Particulate Systems and the Institute of Ruđer Bo{\v s}kovi{\'c} support funds. 


\end{document}